\documentclass{aastex63}

\usepackage{graphicx}
\usepackage[T1]{fontenc}

\newcommand{\st}[1]{_\mathrm{#1}}

\begin{document}

\title{A novel analytic atmospheric $T(\tau)$ relation for stellar models}

\author[0000-0002-4773-1017]{Warrick H. Ball}
\affiliation{School of Physics and Astronomy, University of Birmingham, Edgbaston, Birmingham B15 2TT, United Kingdom}
\affiliation{Stellar Astrophysics Centre, Department of Physics and Astronomy, Aarhus University, Ny Munkegade 120, DK-8000 Aarhus C, Denmark}

\begin{abstract}
  Stellar models often use relations between the temperature $T$ and
  optical depth $\tau$ to evaluate the structure of their
  optically-thin outer layers.  We fit a novel analytic function to
  the Hopf function $q(\tau)$ of a radiation-coupled hydrodynamics
  simulation of near-surface convection with solar parameters by
  \citet{trampedach2014a}.  The fit is accurate to within $0.82$ per
  cent for the solar simulation and to within $13$ per cent for all
  the simulations that are not on either the low-temperature or
  low-gravity edges of the grid of simulations.
\end{abstract}

\section*{}

Standard one-dimensional stellar models require a boundary condition
that summarises the transition from their opaque depths to their
observable outer layers.  These are usually provided by either (a) tables of
surface pressure and temperature computed using separate codes that
model the complicated physics of stellar atmospheres or (b) $T(\tau)$
relations that specify the stratification of the temperature $T$ as a function
of the optical depth $\tau$.  These $T(\tau)$ relations follow the general form
\begin{equation}
  \left(\frac{T(\tau)}{T\st{eff}}\right)^4=\frac{3}{4}\left(\tau+q(\tau)\right)
  \label{e:hopf}
\end{equation}
where $T\st{eff}$ is the effective temperature and $q(\tau)$ is a
\emph{Hopf function}.
We denote the optical depth at which $T=T\st{eff}$ by $\tau\st{eff}$.
Hopf functions can be derived theoretically given some assumptions
(e.g. the Eddington model $q(\tau)=2/3$) or fit to data or detailed atmosphere models.
Two common $T(\tau)$ relations, besides the Eddington relation, are
those of \citet{krishna-swamy1966} and analytic fits to Model C by
\citet{valc}, known as VAL-C \citep[see e.g.][]{paxton2013,sonoi2019}.

In the study of solar-like oscillations, the atmosphere's structure
can affect the mode frequencies appreciably and must therefore be
included in the equilibrium stellar models.  This can be done 
by integrating the equations of the atmosphere's structure and
appending the integrated structure to the interior model.  Alternatively,
one can modify the stellar structure equations and extend the model
to smaller optical depths.  Specifically, one
multiplies the radiative temperature gradient
$\nabla\st{rad}=(\partial\ln T/\partial\ln P)\st{rad}$
by $1+\mathrm{d}q/\mathrm{d}\tau$
\citep[see e.g.][]{mosumgaard2018}.  The first method requires the Hopf function;
the second requires its gradient.

\citet{trampedach2014a} presented Hopf functions for a set of
three-dimensional radiation-coupled hydrodynamics (3D RHD) simulations of
near-surface convection, as well as routines that allow stellar
modellers to interpolate the Hopf functions as a function of the
surface gravity $\log g$ and effective temperature $T\st{eff}$.  The
simulated Hopf functions are most similar to that of VAL-C but do not
tend to a constant temperature at small optical depths.
While we encourage modellers to use the routines to interpolate
in the full suite of Hopf functions, we
present here an analytic function that allows relatively simple
implementation of the solar Hopf function, which is itself somewhat
representative of all the Hopf functions in the grid of models.

The gradient of the Hopf function for the simulation
with solar parameters ($T\st{eff}=5775\,\mathrm{K}$, $\log g=4.438$)
can be approximated by the function
\begin{equation}
  \frac{\mathrm{d}q}{\mathrm{d}x}=\frac{c_1+e^\frac{x-a}{v}}{1+e^\frac{x-b}{w}}\mathrm{,}
  \label{e:dq_dx}
\end{equation}
where $x=\log_{10}\tau$.  This motivates fitting the Hopf function using the integral
of eq.~\ref{e:dq_dx} \citep{wolfram_atm_rnaas2021},
\begin{equation}
  q(x)=c_0 + c_1\left(x-w\ln\left(e^\frac{b}{w}+e^\frac{x}{w}\right)\right)
       + v\,e^\frac{x-a}{v}\,{}_2F_1\left(1,\frac{w}{v};1+\frac{w}{v}; -e^\frac{x-b}{w}\right)\mathrm{,}
  \label{e:q}
\end{equation}
where ${}_2F_1$ is the hypergeometric function.  We found best-fitting parameters $c_0=0.6887302$,
$c_1=0.0668698$, $a=0.9262126$, $b=0.1148743$, $v=0.7657857$ and $w=0.0514999$,
with which the fit reproduces the data to within
$0.82$ per cent over the full range $-4.5<x<2.0$.  The fit is also fairly representative
of all the simulations away from the low-temperature and low-gravity edges of
the grid.  If we exclude simulations with $T\st{eff}<4400\,\mathrm{K}$
or $\log g < T\st{eff}/1000\,\mathrm{K} - 2.2$,
the fit reproduces all the remaining Hopf functions within $13$ per cent.

The hypergeometric function in eq.~\ref{e:q} is not always practical but the term that
contains it does not contribute to the function for $x\lesssim0$.  Ignoring this term
is equivalent to ignoring the denominator in eq.~\ref{e:dq_dx}, in which case the integral
is
\begin{equation}
  q(x)=c_0+c_1(x-b)+v\,e^\frac{x-a}{v}\mathrm{.}
  \label{e:q_approx}
\end{equation}
This is also accurate to within $0.82$ per cent up to $x=0.0741$ for the solar model
and to the
same accuracy up to $\tau\st{eff}=0.5147929$.  Thus, if integrating an
atmosphere using $q(x)$, where one usually terminates at or below
$\tau\st{eff}$, the approximate formula in eq.~\ref{e:q_approx} can be
used.  If including the atmosphere's structure by modifying the
structure equations, then the full equation of the gradient
(eq.~\ref{e:dq_dx}) must be used because we require
$\mathrm{d}q/\mathrm{d}\tau\to0$ for $\tau\gg1$.

\begin{figure*}
  \centering
  \includegraphics[width=\columnwidth]{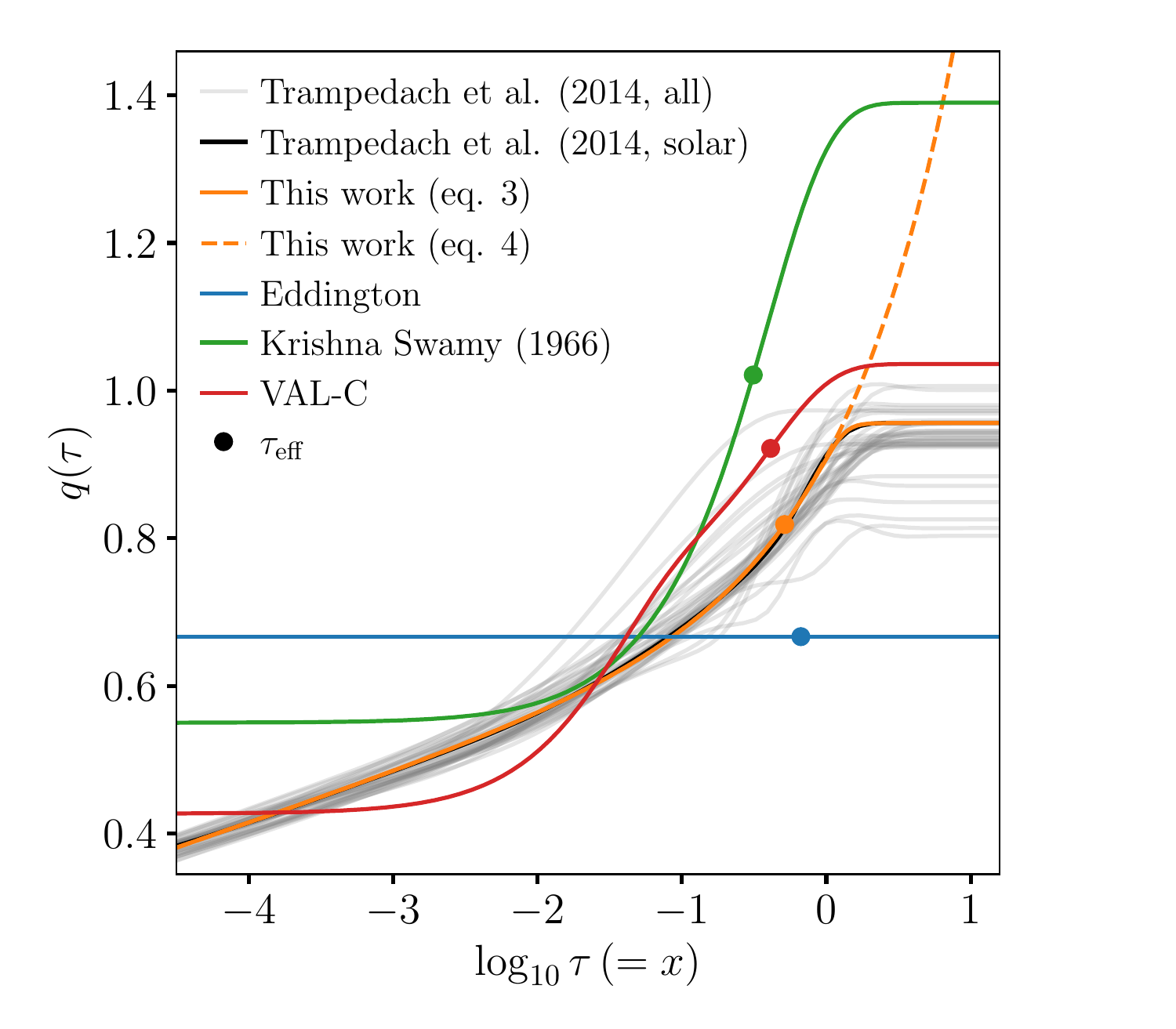}
  \caption{All the Hopf functions $q(x)$ discussed, with the
    corresponding values of $\tau\st{eff}$ indicated by filled circles.  The
    full set of simulations by \citet{trampedach2014a} is shown in
    light grey and their solar simulation in black.  The solid orange
    line is our fit of eq.~\ref{e:q} to the data for the solar simulation
    and the dashed orange line the
    approximate function in eq.~\ref{e:q_approx} using the same
    parameters.  The blue, green and red lines are the popular Hopf
    functions for the standard Eddington atmosphere, the relation by
    \citet{krishna-swamy1966} and a fit to VAL-C
    \citep{valc,paxton2013,sonoi2019}.}
  \label{f:fit}
\end{figure*}

Fig.~\ref{f:fit} shows the Hopf functions we have discussed: the data
from \citet{trampedach2014a}, our fits of eqs~\ref{e:q} and
\ref{e:q_approx} to their solar simulation, and the three widely-used
$T(\tau)$ relations.  Fig.~1 and most of the preceding analysis can be
generated by a publicly available Python
script\footnote{\url{https://github.com/warrickball/atm_rnaas2021}}
\citep{zenodo_atm_rnaas2021}.

\acknowledgements

WHB thanks the UK Science and Technology Facilities Council
(STFC) for support under grant ST/R0023297/1.
Funding for the Stellar Astrophysics Centre is provided by The
Danish National Research Foundation (Grant agreement no.: DNRF106).

\software{NumPy\footnote{\url{http://www.numpy.org}} \citep{numpy2020},
  SciPy\footnote{\url{http://www.scipy.org}} \citep{scipy2020},
  Matplotlib\footnote{\url{http://matplotlib.org}} \citep{matplotlib}
}


\end{document}